\documentclass[aps,prb,floatfix,showpacs,preprint,superscriptaddress]{revtex4}

\usepackage{color}
\usepackage{bm}
\usepackage{amssymb}
\usepackage{amsmath}
\usepackage{amstext}
\usepackage{graphicx}
\usepackage{graphics}
\usepackage{epsfig}
\usepackage{placeins}
\usepackage{psfrag}
\usepackage{subfigure}
\usepackage{ulem}
\usepackage[english]{babel}

\begin{document}

\preprint{ Version March 10, 2008}

\title{Current- and field-driven magnetic antivortices}

\author{Andr\'{e} Drews}
\affiliation{Institut f\"ur Angewandte Physik und Zentrum f\"ur
Mikrostrukturforschung, Universit\"at Hamburg,
  Jungiusstr. 11, 20355 Hamburg, Germany}
\author{Benjamin Kr\"uger}
\affiliation{I. Institut f\"ur Theoretische Physik, Universit\"at Hamburg,
  Jungiusstr. 9, 20355 Hamburg, Germany}
\author{Markus Bolte}
\author{Guido Meier}
\affiliation{Institut f\"ur Angewandte Physik und Zentrum f\"ur
Mikrostrukturforschung, Universit\"at Hamburg,
  Jungiusstr. 11, 20355 Hamburg, Germany}

\date{\today}

\begin{abstract}
Antivortices in ferromagnetic thin-film elements are in-plane magnetization configurations with a core pointing perpendicular to the plane. 
By using micromagnetic simulations, we find that magnetic antivortices gyrate on elliptical orbits similar to magnetic vortices when they are excited by alternating magnetic fields or by spin-polarized currents. The phase between high-frequency excitation and antivortex gyration is investigated. 
In case of excitation by spin-polarized currents the phase is determined by the polarization of the antivortex, while for excitation by magnetic fields the phase depends on the polarization as well as on the in-plane magnetization. Simultaneous excitation by a current and a magnetic field can lead to a maximum enhancement or to an entire suppression of the amplitude of the core gyration, depending on the angle between excitation and in-plane magnetization. This variation of the amplitude can be used to experimentally distinguish between spin-torque and Oersted-field driven motion of an antivortex core.
\end{abstract}

\date{\today}

\pacs{75.60.Ch, 72.25.Ba, 76.50.+g}

\maketitle

\section{Introduction}
Magnetic vortices and antivortices exist in ferromagnetic thin-film elements, where the interplay of demagnetization and exchange energy forces the magnetization out of plane to form a core in the centers.\cite{a1,a18} The orientation of the vortex or antivortex core, denoted as the polarization $p$, is highly interesting for technical applications, e.g. magnetic memory devices, as it can be binary-coded.\cite{a3,a2} Magnetic vortices have been studied intensively in the last years. It has been shown that a vortex core is deflected from its equilibrium position when excited by magnetic fields or spin-polarized currents.\cite{a4,a19} The deflection causes a magnetic stray field which in turn exerts a force on the core.\cite{a28,a29} The resulting gyroscopic motion can be described by a damped two-dimensional harmonic oscillator.\cite{a22} 

\begin{figure}
\includegraphics[width=1\columnwidth,angle=0]{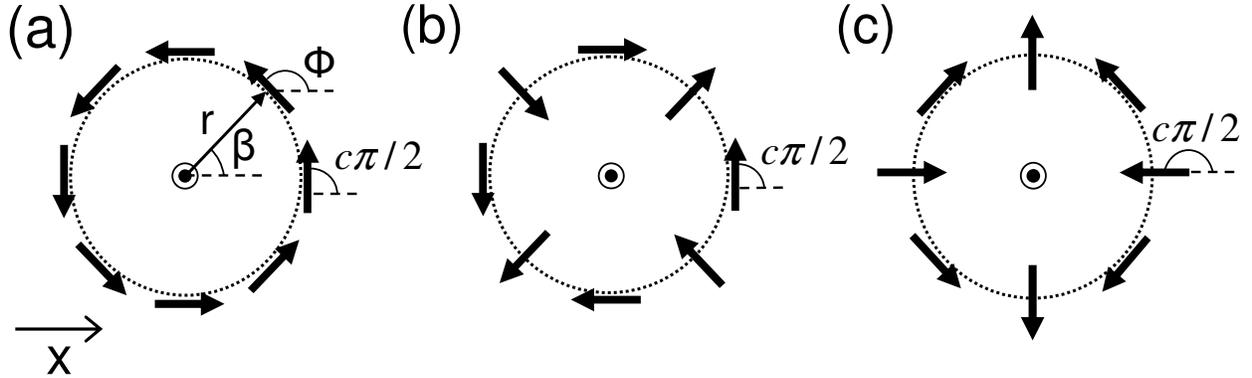}
\caption{\label{Fig1} Definition of $c=2/\pi\cdot(\phi-n\beta)$ for vortices and antivortices by Eq. (1). (a) Magnetic vortex ($n=1$) with $c=1$. (b) Antivortex $(n=-1)$ with $c=1$. (c) Antivortex ($n=-1$) with $c=2$.}
\end{figure}

The dynamics of magnetic antivortices has hithertofore not been studied as intensively as magnetic vortex dynamics. Antivortices appear, e.g., in cross-tie domain walls and individual antivortices have been found in clover-shaped samples.\cite{a7,a30,a3} As illustrated in Fig.~\ref{Fig1}, their in-plane magnetization shows a twofold rotational symmetry that is different from the continuous rotational symmetry of a vortex state. Due to their different  in-plane magnetizations, antivortex dynamics differs from vortex dynamics as is shown in this paper. An understanding of the dynamics of both, antivortices and vortices, is crucial for the description of vortex-antivortex creation and annihilation. These processes have recently received a lot of attention as they are predominant features in the motion of cross-tie walls and in the switching of vortex cores.\cite{a3,a24,a25,a31,a34}
\\
Here we investigate the dynamics of antivortex cores, i.e. sense, phase, and amplitude of gyration, and compare them to the dynamics of magnetic vortices. We show that the direction of the in-plane magnetization around the (anti-)vortex core determines the phase between the exciting alternating magnetic field and the deflection of the (anti-)vortex core. For spin-polarized alternating currents the direction of the in-plane magnetization has no effect on the phase. Both micromagnetic simulations and an analytical model show that simultaneous excitation by magnetic fields and spin-polarized currents can lead to an enhancement or to an entire suppression of the antivortex core displacement.

To classify vortices and antivortices the in-plane magnetization can be described by the relation \cite{a9} 
\begin{equation} \label{eqn1}
\phi=n\beta +\phi_{0}
\end{equation}
between the angular coordinate of the local in-plane magnetization $\phi$ and the angle $\beta$ in real space with respect to the center of the (anti-)vortex core, as shown in Fig.~\ref{Fig1}. The angles $\phi$ and $\beta$ follow the mathematical sense of rotation. For a vortex $n=1$ so that the in-plane magnetization turns in the same direction as the angle in real space with a constant difference $\phi_{0}$ between $\phi$ and $\beta$. For vortices, the angle $\phi_{0}$ is independent of the choice of the axis to which $\beta$ and $\phi$ are measured. Thus for vortices $\phi_{0}$ is an intrinsic quantity which can be expressed by the chirality $c$ as $\phi_{0}=c \pi/2$. In standard geometries and ferromagnetic materials stable vortices can only possess the chiralities $c=1$ or $c=-1$. They can be mapped onto each other by mirroring the sample. In case of an antivortex $n=-1$. This means that the in-plane magnetization turns opposite to the angle in real space.  Though for antivortices the angle $\phi_{0}$ is generally not conserved, because rotations of the sample lead to different values, we define a quantity $c=2\phi_{0}/\pi$ for antivortices with respect to a distinct axis. \footnote{Here, we use the x-axis as distinct axis.} 
Antivortices exhibit values $c$ in the interval $(-2,2]$. A rotation of the antivortex by an anlge of $\Theta$ leads to a change of the c-value of $c=2\Theta$. This is due to the two-fold symmetry of the in-plane magnetization of an antivortex.

\section{Micromagnetic simulations}
To simulate magnetic-field induced antivortex dynamics the OOMMF\cite{a23} code sped up by higher order Runge-Kutta
algorithms is used. The extended code includes the spin-torque terms in the Landau-Lifshitz-Gilbert equation as given by Zhang and Li\cite{a14,a15},
\begin{equation}\label{eqn2}
\begin{split}
{d\boldsymbol{M}\over dt}=&-\gamma '\boldsymbol{M}\times \left ( \boldsymbol{H_{\mbox{eff}}}+{\alpha\over M_{S}}\boldsymbol{M}\times \boldsymbol{H_{\mbox{eff}}}\right )
\\&-(1+\alpha\xi){b_{j}'\over M_{s}^2}\boldsymbol{M}\times \left (\boldsymbol{M}\times(\boldsymbol{j}\cdot \nabla)\boldsymbol{M}\right )
\\&-(\xi-\alpha){b_{j}'\over M_{s}}\boldsymbol{M}\times(\boldsymbol{j}\cdot \nabla)\boldsymbol{M}.
\end{split}
\end{equation}
\begin{figure}
\centering
\includegraphics[width=0.5\columnwidth,angle=0]{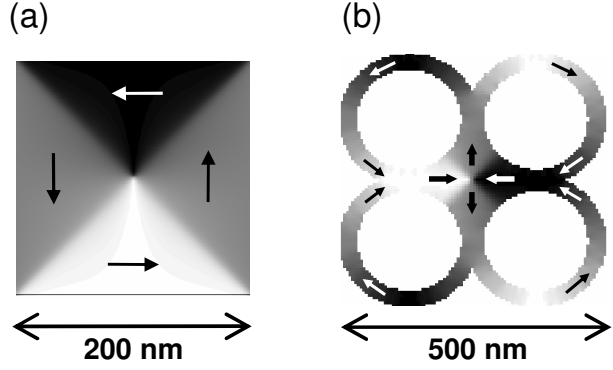}
\caption{Size and shape of (a) the vortex and (b) the antivortex sample.}
\end{figure}
In this equation $\gamma'=\gamma/(1+\alpha^{2})$, where $\gamma$ is the gyromagnetic ratio, $\alpha$ the Gilbert damping, and $\xi$ the ratio between exchange and spin-flip relaxation time. The coupling between local current {\bf j} and magnetization {\bf M} is represented by $b_{j}'=\mu_{B} P/[eM_{s}(1+\alpha^2)]$, where $P$ is the spin polarization. We simulate the excitation of a vortex in a $200\times 200 \times 20~\mbox{nm}^{3}$ permalloy square and an antivortex in a $500 \times 500 \times 40~\mbox{nm}^{3}$ clover-shaped sample. The two geometries are shown in Fig.~2. Different thicknesses of $t=20 ~\mbox{nm}$  for the vortex and $t=40 ~\mbox{nm}$ for the antivortex sample are chosen in order to obtain similar eigenfrequencies for the two geometries. 
We assume a saturation magnetization $M_{s}=8.6\cdot 10^{5}\mbox{~A/m}$, an exchange constant $A=1.3\cdot 10^{-11}~\mbox{J/m}$, a Gilbert damping parameter $\alpha = 0.01$, and a ratio $\xi = 0.9\alpha$ between exchange and spin-flip relaxation time.\cite{a16,a27} A lateral cell size of $4~\mbox{nm}$ is used. Thus the cell size is below the exchange length of permalloy of $l_{ex}=\sqrt{2A/\mu_{0}M_{s}^{2}} \approx 5.3\mbox{~nm}$. The position of the core is defined as the position of the maximum out-of-plane magnetization. To increase the spatial resolution, the magnetization of adjacent cells is matched by a polynomial of second order.\cite{a22}

\begin{figure}
\centering
\includegraphics[width=0.65\columnwidth,angle=0]{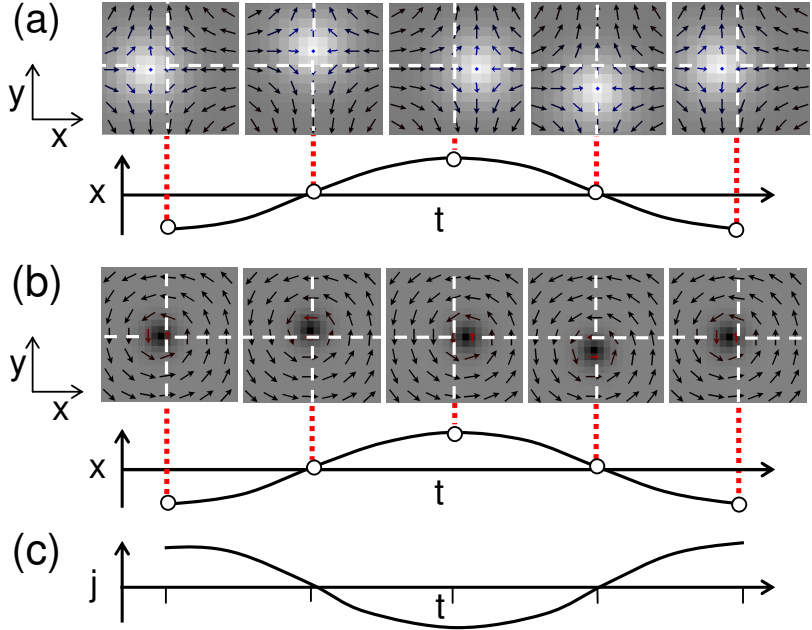}
\caption {Simulation of one gyration period of (a) an antivortex and (b) a vortex. Both have the topological charge $q=-1/2$ and are excited at the resonance frequency (727 MHz for the antivortex and 700 MHz for the vortex) by a current of amplitude $j\cdot P=1.5\cdot 10^{10}~\mbox{A/m}^2$. The graphs below the magnetization images show the deflection in $x$-direction. (c) Exciting alternating current.}
\end{figure}
\section{Excitation by magnetic field and spin-polarized current}
The eigenfrequencies of vortex and antivortex are determined by exciting the core with a current. The free relaxation of the magnetization yields the free frequency $\omega_{f}$ and damping $\Gamma$ of the vortex or antivortex. Because the damping is small compared to the free frequency ($\Gamma\ll\omega_{f}$), antivortex and vortex are weakly damped systems. Thus the frequency of the free oscillation $\omega_{f}$ and the resonance frequency $\omega_{r}$ are approximately the same. The simulated gyration of an antivortex core [$p=1$, see Fig.~3(a)] and a vortex core [$p=-1$, see Fig.~3(b)] both driven by an ac current of amplitude $j\cdot P=1.5\cdot 10^{10}~\mbox{A/m}^2$ are shown in Fig.~3. They both possess the same sense of gyration as defined by the topological charge\cite{a8,a11,a10} $q=np/2=-1/2$. It is known that vortices gyrate counterclockwise with positive and clockwise with negative polarization. \cite{a4} Antivortices, on the other hand, gyrate clockwise with positive and counterclockwise with negative polarization. This is directly observed in the simulations. For small current or small magnetic field amplitudes, the simulated displacement of the antivortex is found to increase linearly with increasing excitation amplitude. This is due to the harmonic potential of the domains' stray field for small displacements of the antivortex core. Thus there is a linear restoring force on the antivortex, which has also been found for a vortex.\cite{a22}
\begin{figure}
\includegraphics[width=0.67\columnwidth,angle=0]{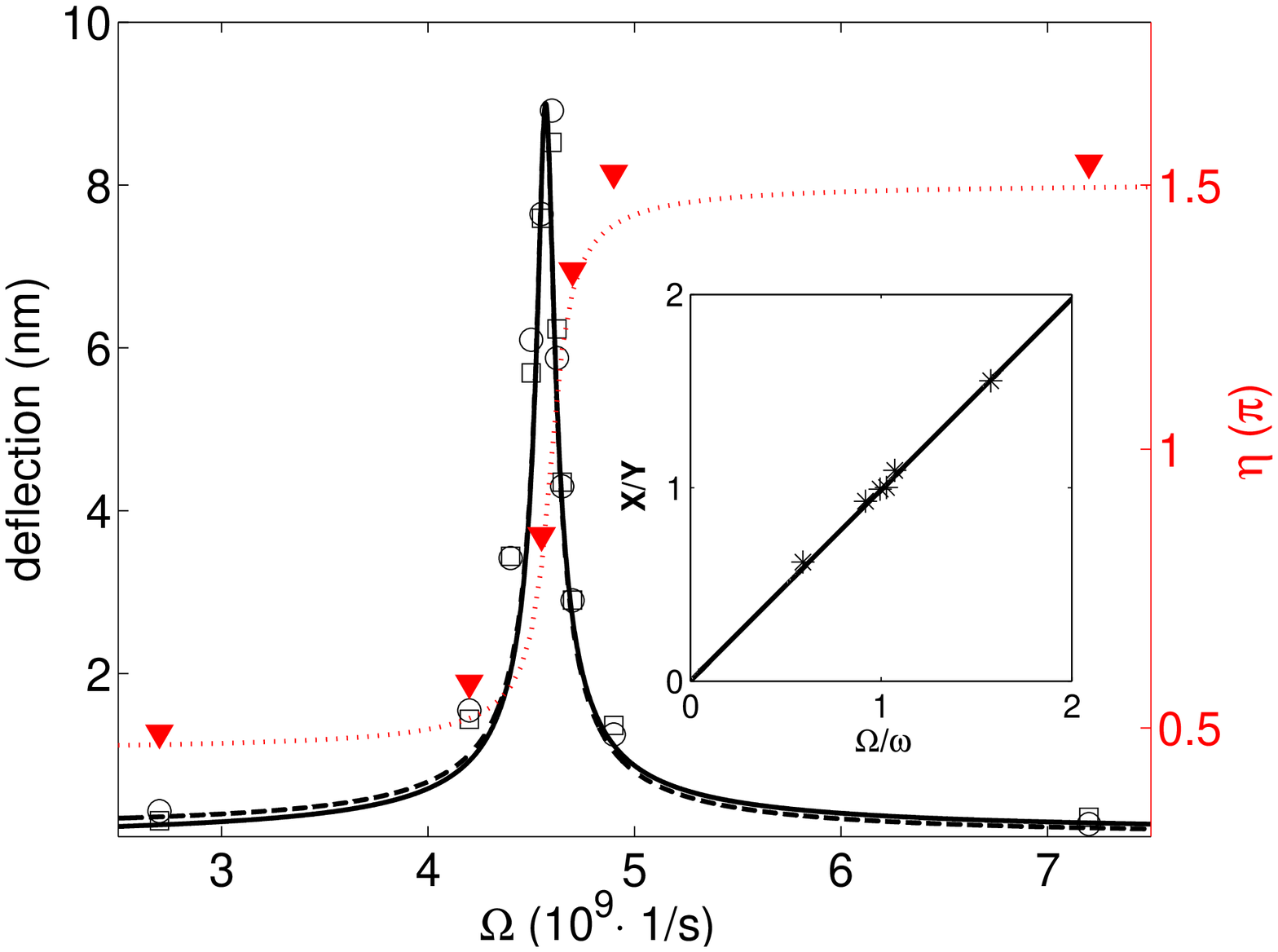}
\includegraphics[width=0.67\columnwidth,angle=0]{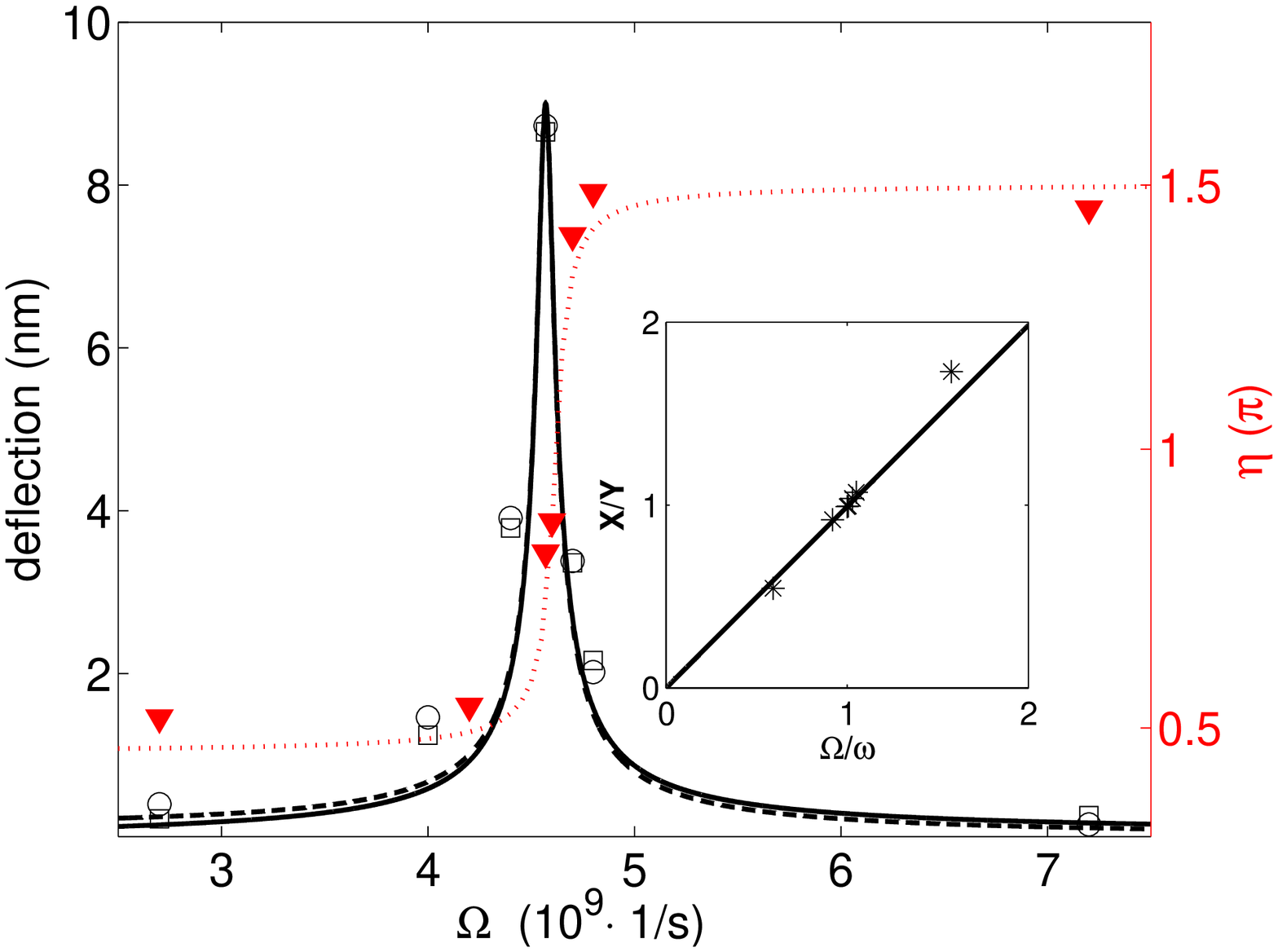}
\caption{ \label{4a}\label{4b}(Color online) Exemplary resonance curves for the semiaxes of the elliptical trajectories and the phase $\eta$ of an antivortex core gyration excited by (a) a current of amplitude $j\cdot P=1.5 \times 10^{10}~\mbox{A/m}^2$ ($c=2$) and (b) a magnetic field of amplitude $H=300~\mbox{A/m}$ ($c=0$). The symbols are results from micromagnetic simulations. The open squares illustrate the semiaxes in $x$-direction, the open circles the semiaxes in $y$-direction. The triangles show the phase $\eta$. The asterisks illustrate the ratio between the semiaxes $x$ and $y$. The solid line is the x-component and the dashed line is the y-component of the amplitude of a fitted resonance curve of a harmonic oscillator. The dotted red line is a fit of the phase $\eta$. The insets show fits of the ratio between the semiaxes $x$ and $y$ as a function of the frequency.}
\end{figure}

To obtain the resonance curve for the amplitude and the phase $\eta$ of the antivortex core, either sinusoidal currents or magnetic fields of frequencies at, above, and below the resonance frequency $\omega_{r}$ are applied. Throughout this paper, the current is applied in $x$-direction while the magnetic field is applied in $y$-direction. The resonance curve of a harmonic oscillator\cite{a32} with a resonance frequency $\omega_{r}/2\pi=727~\mbox{MHz}$ and a damping ${\Gamma/2\pi}=6.4~\mbox{MHz}$ matches very well the numerical data, as shown in Fig.~4~(a,b). In general the antivortex gyrates on elliptical orbits.  The semi-major (semi-minor) axis of the ellipses at frequencies below resonance changes into the semi-minor (semi-major) axis at frequencies above resonance. At resonance the trajectories are circular. This is illustrated in Fig.~4~(a) for a current-driven antivortex with $c=2$ and in Fig.~4~(b) for a magnetic-field driven antivortex with $c=0$. For both the semi-major axes point in $y$-direction at frequencies below and in $x$-direction at frequencies above  resonance.\\ 
The phase $\eta$ is defined by the temporal delay between the maximum of the applied current or field and the maximum core displacement in $x$-direction. Like for a harmonic oscillator the phase $\eta$ changes by $\pi$ when the exciting frequency $\Omega$ is increased from values well below to values well above the resonance frequency $\omega_{r}$. This is illustrated in Fig.~4 (a) for current and in Fig.~4 (b) for magnetic-field excitation.

\begin{figure}
\includegraphics[width=0.8\columnwidth,angle=0]{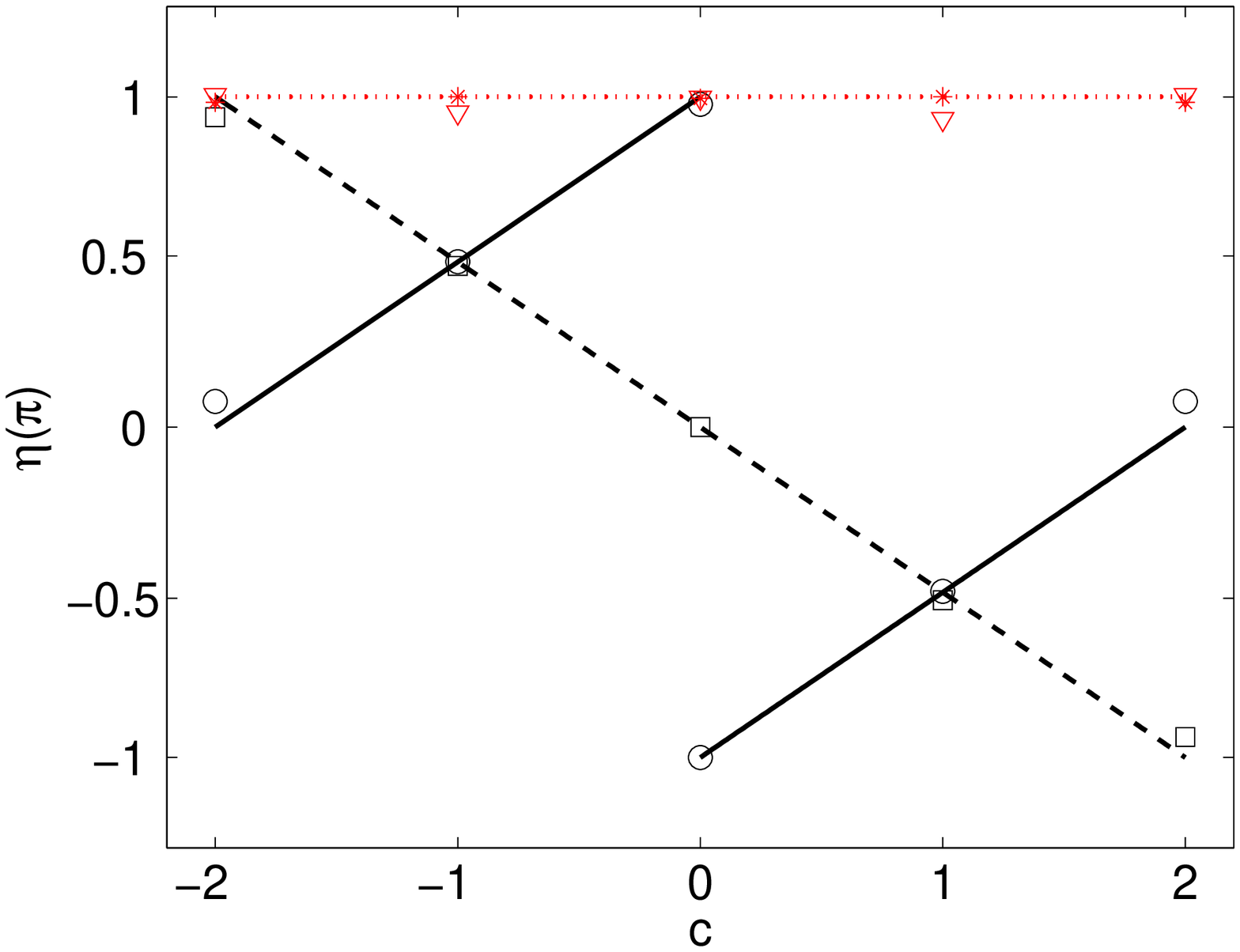}
\caption{\label{Fig5} (Color online) Phase $\eta$ between excitation and displacement for an antivortex core gyration at resonance. The dotted red line illustrates the phase when the antivortex core is excited by a spin-polarized current. The asterisks represent corresponding results from micromagnetic simulations for positive core polarization ($p=1$) and the triangles for negative polarization ($p=-1$). For excitation with a magnetic field the solid line illustrates the phase for positive core polarization ($p=1$), the dashed black line for negative polarization ($p=-1$). The numerical results are depicted by open circles and squares.}
\end{figure}

We numerically simulate the dependence of the phase on the direction of the in-plane magnetization by exciting at resonance antivortices of all possible integer $c$-values and of both polarizations $p=-1$ and $p=1$. For current excitation for all $c$-values the antivortex cores are deflected into the physical current direction ($\eta=\pi$). For magnetic field excitation, the phase is found to depend on the direction of the in-plane magnetization as shown in Fig.~5. For a constant frequency the phase varies by $2\pi$ when $c$ is changed from $-2$ to $2$, i.e. when rotating the sample by $\pi$ with respect to the magnetic field. At resonance the phase changes from $0$ to $2\pi$ for $p=1$ and from $\pi$ to $-\pi$ for $p=-1$ as illustrated in Fig.~5. 

The phase $\eta$ and its dependence on the direction of the in-plane magnetization is studied analytically by using the equation of motion for vortices and antivortices \cite{a32} assuming low damping ($\omega_{r}\gg\Gamma$). The equation for the deflection 
\begin{equation}\label{eqn3}
\begin{pmatrix} 
x \\ y
\end{pmatrix}=
-\chi\cdot 
\biggl (\begin{matrix} 
& v_{H} \sin({\pi c\over {2}})\omega+  \left( v_{H}p\cos({\pi c\over 2})+v_{j}\right) i \Omega\\
&\left(  v_{H}n \cos({\pi c\over 2})+ v_{j}np\right) \omega -v_{H} np\sin({\pi c\over 2 }) i\Omega
\end{matrix}
\biggr )\cdot e^{i\Omega t}
\end{equation}
is derived from the Thiele equation\cite{a29} for excitation with magnetic fields and the extension by Thiaville\cite{a28} for spin-polarized currents. A harmonic potential due to the demagnetizing field is assumed.\cite{a32} The velocity due to the adiabatic spin-torque term is $v_{j}=b_{j}j_{0}$, the velocity due to the magnetic field $v_{H}=\gamma H_{0}l/(2\pi)$, and the susceptibility of a harmonic oscillator is $\chi=1/{[\omega^2+(i\Omega+\Gamma)^2]}$. Equation~(3) states that a change of the $c$-value leads to a rotation of the magnetic force. This in turn causes a dependence of the phase on the in-plane magnetization. For example, a change from $c=-1$ to $c=1$ is equivalent to a rotation of the magnetic force by an angle of $\pi$. 
At resonance Eqn.~(3) yields the deflection
\begin{equation}
x(c,p,t)=\frac{v_{H}}{2\Gamma} e^{i\Omega t}e^{i\pi(\frac{1+p-pc}{2})}.
\end{equation} 
The maximum excitation is reached for $i\Omega t=-i\pi(\frac{1+p-pc}{2})$, so that the phase of the antivortex motion that is induced by a magnetic field at resonance reads 
\begin{equation}
\eta_H=-\pi\biggl (\frac{1+p-pc}{2} \biggr).\label{eqn5}
\end{equation} 
For purely current-driven excitation Eqn.~(3) gives 
\begin{equation}
x(c,p,t)=-\frac{v_{j}}{2\Gamma}e^{i\Omega t}.
\end{equation}
Hence the phase induced by a current is $\eta_{j}=\pi$ 
at resonance. This means that for current excitation the phase is independent of the direction of the in-plane magnetization as well as of the polarization. Figure ~\ref{Fig5} demonstrates that the analytical results agree well with the simulations.

\section{Amplitude variation of gyration }
\begin{figure}
\includegraphics[width=0.7\columnwidth,angle=0]{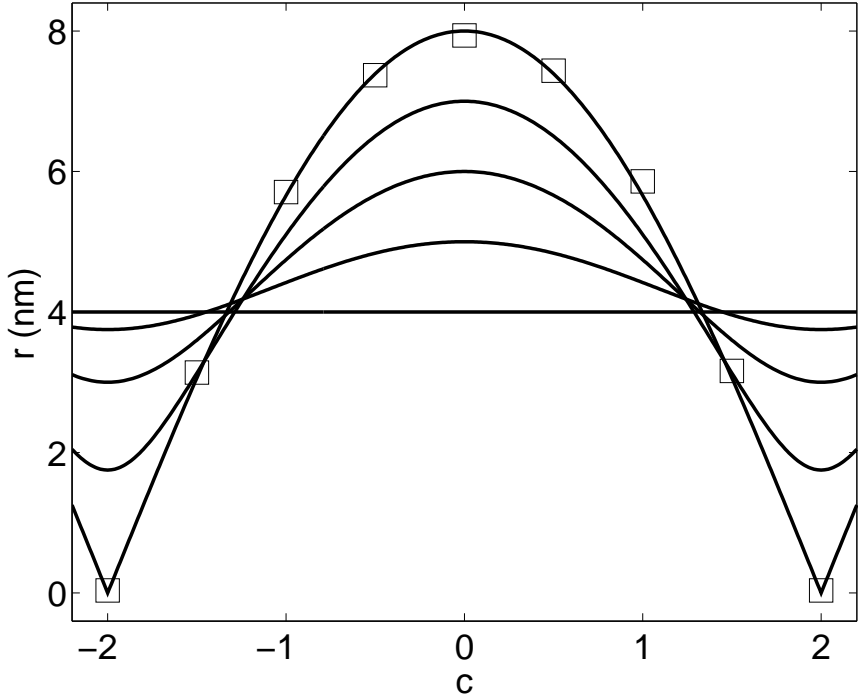}\label{Fig6}
\caption{ Amplitude of displacement of an antivortex core with polarization $p=1$ at resonance. The antivortex is excited simultaneously by a magnetic field in $y$-direction and a current in $x$-direction. The symbols denote simulated results for $v_{H}=v_{j}$, the lines are fits for different ratios $v_{H}/v_{j}$ according to Eqn.~(7) .} 
\end{figure}
In the following we simulate antivortices that are excited simultaneously by a magnetic field and a spin-polarized current. First the antivortex is excited by a current in $x$-direction in the absence of a magnetic field. Then the amplitude of a magnetic field in $y$-direction is tuned until the antivortex core gyration possesses the same amplitude as under current excitation. In our case the core is excited by a spin-polarized current of amplitude $j\cdot P=0.7\cdot 10^{10}~\mbox{A/m}^2$ that corresponds to a magnetic field of amplitude $\mbox{H}=122~\mbox{A/m}$.  
Then the current and the magnetic field are applied simultaneously. Different directions of the in-plane magnetization (see Fig.~6) are chosen to investigate the c-dependent variation of the core amplitude for a positive polarization. The simulation shows a doubling of the amplitude at $c=0$ and a complete suppression at $c=2$. Thus a superposition of the deflection by current and a perpendicular field leads to an amplitude variation in dependence on the direction of the in-plane magnetization of the sample. This is due to the c-dependent phase between antivortex core displacement and magnetic field, see Eqn.~(\ref{eqn5}). The forces due to current and magnetic field are proportional to the deflections. If they are parallel or antiparallel, an enhancement or suppression of the core displacement is found, respectively. When both deflections have the same amplitude, the amplitde of gyration can be doubled or completely quenched, as shown in Fig.~{\ref{Fig6}}.
\\ Using the addition theorem one can derive from Eqn.~(3) the c-dependent amplitude variation\begin{equation}
r=r_{0}\sqrt{\biggl ({{v_{H}}\over {v_{j}}}-p\biggr )^2 \sin^2 \biggl ({{\pi c}\over{4}}\biggr )+\biggl ({{v_{H}}\over {v_{j}}}+p\biggr )^2 \cos^2 \biggl ({{\pi c}\over {4}}\biggr )
}.
\end{equation}
 of the antivortex core gyration at resonance.
This is a general expression for arbitrary ratios $v_{H}/v_{j}$ between the antivortex core velocities due to current and to magnetic field, for both polarizations and all $c$-values. The amplitude is plotted in Fig.~6.

\begin{figure}[t!]
\includegraphics[width=0.7\columnwidth,angle=0]{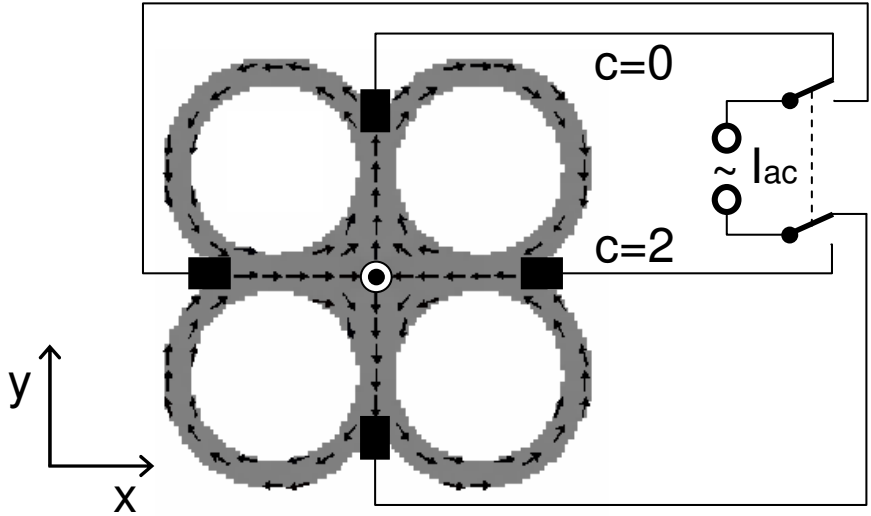}
\caption{\label{Fig8}{Proposed setup with electrical contacts to excite a single antivortex, here with polarization $p=1$. The quantity $c$ depends on the direction of the exciting ac current. For current in $x$- or $y$-direction the deflection is suppressed or amplified, respectively.}}
\end{figure}

An inhomogeneous current distribution in the direction of the film normal generates a non-zero Oersted field perpendicular to the current.\cite{a33}
For the experimental proof of a dependence of the amplitude on the direction of the in-plane magnetization, we propose a setup with a clover-shaped sample that is illustrated in Fig.~\ref{Fig8}. A similar sample was investigated by Shigeto et al. \cite{a7} with magnetic-force microscopy. Excitation by a spin current in $x$-direction or $y$-direction through electrical contacts corresponds to a direction of the in-plane magnetization for $c=2$ and $c=0$, respectively. For the polarization $p=1$ we expect a suppressed motion when the current is applied in $x$-direction and an enhanced amplitude when it is applied in $y$-direction. The variation of the amplitude for $c=0$ and $c=2$ could be used to determine the ratio between the forces on the antivortex core due to an Oersted field and a current.
\section{conclusion}
In conclusion, we have demonstrated by micromagnetic simulations that antivortices excited by spin-polarized ac currents or magnetic fields gyrate on elliptical orbits. These orbits can be well described by the analytical model of a two-dimensional harmonic oscillator. The sense of gyration of antivortices depends solely on the topological charge $q=np/2$. The phase of the antivortex motion excited by an alternating magnetic field depends also on the direction of the in-plane magnetization. Antivortices that are excited simultaneously by a spin-polarized current and a magnetic field show an enhancement or a suppression of the deflection´s amplitude in dependence on the direction of the in-plane magnetization. The effect of the amplitude variation in dependence on the in-plane magnetization can be used to experimentally investigate the influence of Oersted fields in current-induced antivortex dynamics.
\begin{acknowledgments}
We thank Ulrich Merkt and Daniela Pfannkuche for valuable discussions and encouragement. Financial support by the Deutsche Forschungsgemeinschaft via the Graduiertenkolleg 1286 "Functional metal-semiconductor hybrid systems" and via Sonderforschungsbereich 668 "Magnetismus vom Einzelatom zur Nanostruktur" is gratefully acknowledged.
\end{acknowledgments}


\end{document}